\documentclass[aps,prl,showpacs,floatfix,twocolumn,superscriptaddress,
amsmath,amssymb]{revtex4}

\usepackage{graphicx}
\usepackage{hyperref}

\begin{document}
\title{Generation of high-frequency radiation in semiconductor superlattices
with suppressed space-charge instabilities}
\author{Kirill N. Alekseev}
\affiliation{Department of Physical Sciences,
P.O. Box 3000, FIN-90014 University of Oulu, Finland}
\author{Natalia V. Demarina}
\affiliation{Radiophysics Department, Nizhny Novgorod State University,
603950 Nizhni Novgorod, Russia}
\author{Maxim V. Gorkunov}
\altaffiliation[Also at ]{ Institute of Crystallography RAS, Moscow
119333, Russia}

\affiliation{Department of Physical Sciences,
P.O. Box 3000, FIN-90014 University of Oulu, Finland}

\pacs{73.21.Cd, 07.57.Hm, 72.20.Ht}

%
\begin{abstract}
We theoretically investigated the scheme allowing to avoid
destructive space-charge instabilities and to obtain a strong gain
at microwave and THz frequencies in semiconductor superlattice
devices. Superlattice is subjected to a microwave field and a
generation is achieved at some odd harmonics of the pump
frequency. Gain arises because of parametric amplification seeded
by harmonic generation. Negative differential conductance (NDC) is
not a necessary condition for the generation. For the mode of
operation with NDC, a limited space-charge accumulation does not
sufficiently reduce the gain.
\end{abstract}
\maketitle
There exists a strong demand for miniature, solid-state, room
temperature operating sources and detectors of THz radiation
($0.2$-$10$ THz). The need is caused by a rapid progress of THz
sciences and technologies ranging from the astronomy to the biosecurity.
Semiconductor superlattices (SLs) \cite{esaki70}, operating in the
miniband transport regime, are interesting electronic devises
demonstrating properties of both nonlinear and active media.
Nonlinearity of voltage-current (UI) characteristic of SL gives rise
to a generation of
harmonics of microwave and THz radiation \cite{esaki71,ignatov76};
for a high electron mobility and a high doping, even such strongly
nonlinear phenomena as chaos \cite{alekseev96} and symmetry
breaking \cite{alekseev98} have been predicted. On the other hand,
Bloch oscillations of electrons within a miniband of SL cause an
appearance of negative differential conductance (NDC)
for dc fields (voltages) larger than the critical Esaki-Tsu
$E_c$ \cite{esaki70}. Remarkably, the static NDC is
accompanied by a small-signal gain for ac fields of very broad
frequency range from zero up to several THz \cite{kss}. The feasible device,
employing such active media properties of SL,
is known as the THz Bloch oscillator. However, an existence of gain at
low frequencies causes space-charge instabilities and a formation of
high-field domains inside SL \cite{kss,buettiker77}. These domains
are believed to be distractive for the Bloch gain. Recent experiments
demonstrate that even a use of sophisticated design of SL device
\cite{savvidis} or a development of delicate spectroscopy
techniques \cite{lisauskas} enable to observe only a very weak
Bloch gain in dc-biased SLs.
\par
In these respects important questions arise: Is it possible to
obtain a gain for a high-frequency field employing only a nonlinearity
of UI-characteristic of SL? Can we avoid formation of
domains in the schemes based on nonlinear mechanisms of gain?
\par
In the present Letter, we examine a feasibility of high-frequency superlattice
oscillator supporting the regimes either with strongly suppressed
or even without space-charge instabilities.
We analyze the situation when a pure ac pump field of microwave
frequency range is applied to SL. Generation arises at third, seventh
\textit{etc.} harmonics of the pump. In the weak-probe field
limit, the mechanism for a gain consists in a combination of parametric
amplification of the probe seeded by a frequency multiplication of
the pump, as well as of nonparametric absorption. Remarkably,
the oscillator can operate even if the amplitude of pump
field does not reach $E_c$ preventing an instability of charge
distributions inside SL.
Thus, we demonstrate the possibility to obtain a gain for a
high-frequency field employing only nonlinear characteristics of
SL without using its active medium properties. Efficiency of the oscillator
increases if SL switches partly to NDC. For the regime of operation with NDC, we
clearly show that a space-charge accumulation can only
limit but not destroy the gain.
This mechanism of space-charge instability stabilization
is rather general and applicable to other
SL devices, like the Bloch oscillator.
\par
Part of the presented here results has been reported at the conferences
\cite{alekseev_conf}.
\par
The paper is organized as follows. (1) Analytic consideration of
a small-signal gain; (2) 1D and 3D numerical study of a large-signal
gain; (3) Effect of a finite $Q$ of resonator on the oscillator efficiency;
(4) Analytic analysis of space-charge wave stability in the linear
approximation and within the framework of drift-diffusion model; (5) Numerical
analysis of nonlinear stage of the instability within the
drift-diffusion model; (6) Direct comparison of the gain found in
spatially-homogeneous approximation with the gain computed with
an account of space-charge effects: Calculations using both the 1D drift-diffusion
model and the 3D ensemble Monte Carlo method; (7) Effects arising beyond the quasistatic
approximation; (8) Brief historical review.
Some details of the calculations are presented in the separate appendixes.
\par
We consider the response of miniband
electrons to the action of ac field $E(t)=E_{p}+E_{pr}$, where
$E_{p}=E_{\omega}\cos\omega t$ is the strong pump
and the probe, $E_{pr}=E_n\cos\omega_n t$,
is the $n$th odd harmonic of the pump, $\omega_n=n\omega$ ($n=3,5,7,\ldots$).
SL is placed in a cavity providing a
feedback only for the field with frequency $\omega_n$.
In the quasistatic limit $\omega\tau<1$, $\omega_n\tau<1$, the dependence
of the drift electron velocity $V$ on the electric field can be well
described by the Esaki-Tsu formula
$V(E)=2 V_p(E/E_c)/[1+(E/E_c)^2]$,
where $E_c=\hbar/e a\tau$ is the critical field ($a$ is the SL period,
$\tau$ is the intraminiband relaxation time)
and $V_p\equiv V(E_c)$ is the peak electron velocity
\cite{esaki70}. We suppose an operation at room temperature.
The absorption of the probe field in SL is defined in
scaled units as
$A=\langle v(t)\cos(\omega_n t) \rangle$,
where $v=V(E)/V_p$ and averaging $\langle\ldots\rangle$
is performed over the period $T=2\pi/\omega$.
\par
First, we are interested in a small-signal gain in SL ($E_n\ll
E_{\omega}$). Expanding $v(E)\approx
v(E_{p})+v^{\prime}(E_{p})\times E_{pr}$ (prime means the
derivation with respect to $E$), we present the total absorption
$A$ as the sum of $A_{h}=\langle v(E_{p})\cos(n\omega t)\rangle$,
$A_{coh}=\langle v^{\prime}(E_{p})\cos(2n\omega t)\rangle E_n/2$
and $A_{inc}=\langle v^{\prime}(E_{p})\rangle E_n/2$ terms, which
can be called the harmonic, the coherent and the incoherent
absorption components, correspondingly (for details see
Appendix~A).
\par
We see that while $A_{coh}$ and $A_{inc}$ are dependent on both
$E_{\omega}$ and $E_n$, the term $A_{h}$ is a function of only
pump strength $E_{\omega}$, and therefore $A_{h}$ gives the main
contribution to the absorption of a weak probe. Substituting the
Esaki-Tsu dependence in the definition of $A_{h}$, we find
$A_{h}=(-1)^k [2( b-1)^{2k+1}]/b{\bar E}_{\omega}^{2k+1}$, where
$b=(1+{\bar E}_{\omega}^2)^{1/2}$, ${\bar E}_{\omega}\equiv
E_{\omega}/E_c$ and $2k+1=n$ ($k=1,2,3\ldots$). This equation
describes odd harmonics of the current (\textit{cf} Eq. 17 in
\cite{ignatov76}); in the limit of weak pump ${\bar E}_{\omega}\ll
1$ it takes familiar form $\propto E^n$ \cite{esaki71}.
Importantly, $A_{h}$ is negative for the odd values of $k$. Therefore, generated
harmonics with $n=3,7,\ldots$ can provide \textit{seeding gain} for
an amplification of a probe field. It has no threshold in the
amplitude of pump $E_{\omega}$. Therefore, if the pump amplitude
is less than the critical field, $E_{\omega}/E_c<1$, the gain at harmonics
will be not accompanied by the space-charge instabilities in SL.
\par
We turn to the analysis of the term $A_{coh}$ that describes a
parametric amplification of the probe field due to a coherent
interaction of the pump and the probe fields in SL. Numerical calculation
of the integral demonstrates that $A_{coh}$ is always negative for
odd $n=3,5,7$. In the limit of weak pump ${\bar E}_{\omega}\ll 1$,
we find $A_{coh}\propto {\bar E}_{\omega}^{2n}{\bar E}_n$. Now we
can describe the amplification of a weak signal at $3\omega$ under
the action of a weak pump. Third harmonic of a weak pump is generated
at cubic nonlinearity of SL characteristic, then it can get a seeding
gain at the same nonlinearity ($A_{h}\propto E_{\omega}^3$), but the
next step of parametric gain, $A_{coh}$, uses already the seventh
order ($\simeq E^7$) nonlinearity in the $V(E)$-dependence. The $5$th
harmonic also can be generated in SL. However, in contrast to the
case of the $3$d harmonic, it cannot be further amplified ($A_{h}>0$)
and field in the cavity, tuned to $5\omega$, eventually evolves to
zero.
\par
Finally, nonparametric effects are described by the term
$A_{inc}={\bar E}_n/( 1+{\bar E}_{\omega}^2)^{3/2}$ (${\bar
E}_{n}\equiv E_{n}/E_c$). We should notice that this is the only
term which can exist in the expression for total absorption of SL
in the case $\omega_n/\omega$ is not an integer. The absorption
$A_{inc}$ is always positive. We will see that the absorption due
to the incoherent interactions of fields plays an important role
in the stabilization of space-charge instability in SL.
\begin{figure}[htbp!]
\includegraphics[width=0.7\linewidth]{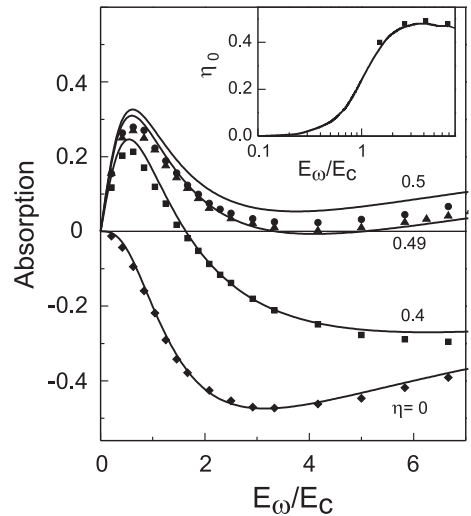}
\caption{ \label{fig1} The dependence of absorption at $3\omega$
on the pump amplitude $E_{\omega}$ for the different
relative probes, $\eta$. Inset:
Maximal relative amplitude of the probe possessing gain,
$\eta_0$, as function of the pump $E_{\omega}$.
Calculation with Esaki-Tsu
characteristic (solid) and Monte Carlo technique (symbols).}
\end{figure}
\par
To find the efficiency of oscillator we need to consider the
effects of both a large-signal gain and a finite $Q$-factor of
resonator. The stationary value of the cavity field, $E_n^{st}$,
is determined by the balance of gain and loss \cite{lasers} as
${\bar E}_n^{st}/Q^{\prime}=-A( {\bar E}_{\omega},{\bar
E}_n^{st})$, where $Q^{\prime}=(\omega_{pl}^2\tau/\omega_n) Q$,
${\bar E}_n^{st}=E_n^{st}/E_c$ and $\omega_{pl}=(4\pi e^2 N/
\epsilon m_0)^{1/2}$ is the miniband plasma frequency ($N$ is the
doping density, $m_0$ is the electron mass at the bottom of
miniband and $\epsilon$ is the averaged dielectric constant of
SL), (for details see Appendix B). We found a large-signal
absorption $A(E_{\omega},E_n)$ using two methods: (i) simple 1D
calculations with Esaki-Tsu characteristic and (ii) 3D
single-particle Monte Carlo computations with an account of
electron scattering at optical and acoustic phonons
\cite{demarina}. For the Monte Carlo computations we consider
GaAs/AlAs SL of the period $a=6.22$ nm and the miniband width
$\Delta=24.4$ meV. Static UI-characteristic of this SL can be well
described by the Esaki-Tsu formula with $E_c=4.8$ kV/cm,
$V_p=1.44\times 10^6$ cm/s and $\tau=220$ fs. We took the
frequency of pump field $\omega=100$ GHz. These are our default
parameters for all computations.
\par
We start with the case of an ideal cavity ($Q\rightarrow\infty$). The
dependence $A(E_{\omega})$ for $3\omega$-generation and for
the different values of relative probe amplitude,
$\eta=E_3/E_{\omega}$, are shown in Fig.~\ref{fig1}. Results of
simple 1D theory and 3D Monte Carlo simulations are in a good
agreement. For $\eta=0$ and for small $\eta\ll 1$, the dependence of
$A({\bar E}_{\omega})$ follows to the corresponding dependencies for the seeding gain,
$A_h({\bar E}_{\omega})$, and the small-signal gain. With a
further increase of $\eta$, the gain
decreases and finally the absorption becomes zero for some
$\eta=\eta_0$: $A(E_{\omega},\eta_0)=0$.
The value $\eta_0^2=[E_n^{st}/E_\omega]^2$
determines the maximal SL
oscillator efficiency for the given pump. Inset in Fig.~\ref{fig1}
shows the dependence of $\eta_0$ on the pump amplitude. If the
amplitude of pump does not reach NDC, $E_{\omega}<E_c$, the
maximal oscillator efficiency is less than 5\%. For the
oscillations in NDC regime, $\eta_0^2$ reaches $23$\% for
$E_\omega/E_c\approx 4$. The maximal efficiency for
$7\omega$-oscillations is $5$\% ($E_\omega/E_c\approx 5$).
\begin{figure}[htbp!]
\includegraphics[width=0.7\linewidth]{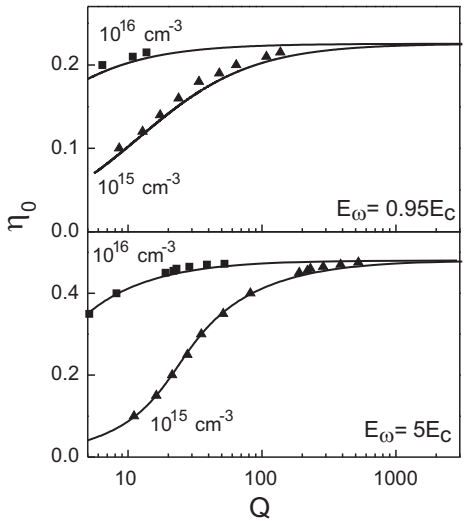}
\caption{ \label{fig2} Dependence of the stationary relative amplitude
of oscillations $\eta_0$ on $Q$ for the pump amplitudes
$E_{\omega}/E_c=0.95$ (upper) and $E_{\omega}/E_c=5$ (lower).
Calculation using Esaki-Tsu characteristic (solid) and Monte Carlo
technique (symbols).}
\end{figure}
\par
We also evaluated the oscillator efficiency in the case of finite
$Q$. Fig.~\ref{fig2} represents the dependencies of
$\eta_0=E_3^{st}/E_{\omega}$ on $Q$ for the doping densities
$N=10^{15}$ and $10^{16}$ cm$^{-3}$. We see that for a heavy doped
SL the efficiency of generation can reach the maximal efficiency ($\approx 23$\%)
even in a cavity with $Q<100$ for both operational regimes: with
NDC and without NDC (\textit{cf} Figs.~\ref{fig2} and \ref{fig1}).
We should also note that oscillator's start up also does not
require a high-$Q$ resonator because of large small-signal gain in
SL.
\par
Although the oscillator can operate even without NDC, the
maximum of efficiency within spatially-homogeneous approximation
is reached for $E_{\omega}/E_c>1$. Therefore we turn to the
consideration of an evolution of space-charge instabilities in SL.
For $\omega\tau<1$ a space-time evolution of the electron density
$\rho(x,t)$, the current density $j(x,t)$ and the field $E(x,t)$
inside a SL of the total length $L$, driven by the given voltage
$U_p(t)=U_{\omega}\cos\omega t$ ($U_{\omega}=E_{\omega} L$), can
be well described by  the drift-diffusion (DD) model
\cite{ignatov87}. Set of equations consists of the current
equation $j=e \rho V(E)-D(E)\partial \rho/\partial x$, the Poisson
equation $\partial E/\partial x=4\pi\epsilon^{-1}(\rho-N)$, the
relation to the applied voltage $U(t)=\int_0^L E(x,t)dx$, and the
continuity equation $e\partial\rho/\partial t+\partial j/\partial
x=0$ \cite{dd_review}. Following the Einstein relation the
diffusion coefficient is $D=k_B T V /E$. Linearizing these
equations, we find that small fluctuations of
space-charge with long wavelength will grow, if the dielectric
relaxation increment, $\omega_d \int_0^t [\partial V(E_p(t))
/\partial {\bar E}] dt$ (with $\omega_d=\omega_{pl}^2\tau$), is
\textit{negative}. Taking the integral over the period of pump, $t=T$, we see that
the increment is proportional to $A_{inc}$, which is \textit{always
positive} for all $n$ and $E_{\omega}$. Therefore the system is stable against
small fluctuations of charge or field. That is the
limited-space-charge accumulation (LSA) mode of  SL oscillator
operation. However, in contrast to the traditional LSA mode in
Gunn diodes driven by dc and ac voltages \cite{copeland_67}, LSA
in SL does not require a large amplitude of ac field.
\par
We also considered the influence of voltage harmonics calculating numerically
the increment for $U(t)=U_p(t)+U_n\cos\omega_n t$.
For the practically important range $U_n<U_n^{st}$, we found that LSA still works.
Next, solving numerically DD-model, we consider dynamics of large fluctuations.
We find that even large fluctuations of charge-density are resolved  during the time of dielectric
relaxation $\propto\omega_d^{-1}$. Therefore this type of LSA in SL works even in the nonlinear regime.
\par
\begin{figure}[htbp!]
\includegraphics[width=0.9\linewidth]{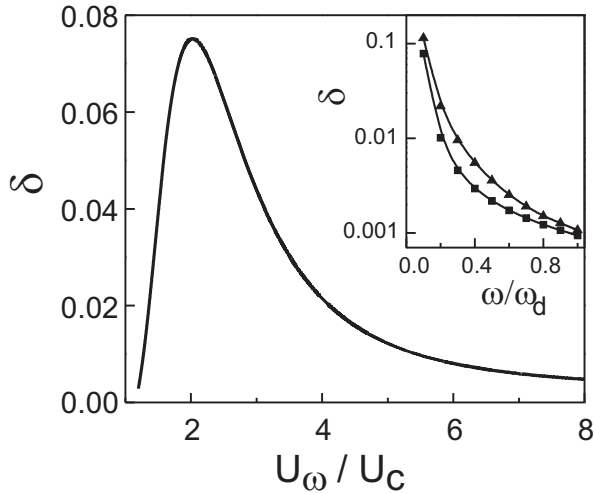}
\caption{
\label{fig3}
LSA regime of SL oscillator.
Relative reduction of gain at $3$d harmonic $\delta$ vs voltage amplitude $U_\omega$
for $\omega/\omega_d=0.1$ and $\eta=U_3/U_\omega=0.1$.
Inset: $\delta$ as a function of scaled frequency $\omega/\omega_d$ for
$U_\omega/U_c=2$, $\eta=0$ (squares) and $\eta=0.4$ (triangles).}
\end{figure}

Finally, we directly calculated the influence of inhomogeneous
distributions of space-charge on a gain in SL. We determined
the absorption at the $n$th harmonic of voltage as
$A_d=\langle\langle(j(x,t)/j_0)\cos\omega_n t\rangle\rangle$,
where $j_0=e V_p N$ and averaging
$\langle\langle\ldots\rangle\rangle$ is performed both over the
period $T$ and the length $L$. In the computations a formation of
domains was caused by a Gaussian spread in the value of $E_c$ (or
$\tau$). Domains were periodically created during the part of
period $T$ when the SL was switched to the state with NDC,
and then they were annihilated during another part of $T$.
We computed the relative decrease in
gain, $\delta=(A-A_d)/A$, for different $\omega$ and $U_\omega$.
Dependence of $\delta$ on $U_\omega/U_c$, for a long SL of $130$
periods and for $\omega/\omega_d=0.1$, is shown in
Fig.~\ref{fig3}. Reduction of gain due to LSA is $2$\% for optimal
$U_\omega/U_c=4$ and it is less than $8$\% overall. For our
default parameters with $\omega=100$ GHz, the value
$\omega/\omega_d=0.1$ corresponds to the doping $N=2\times
10^{16}$ cm$^{-3}$. For the same SL parameters but for the pump
frequency of tens of GHz, the condition $\omega/\omega_d=0.1$ is satisfied for
$N\simeq 10^{15}$ cm$^{-3}$. As is evident from Fig.~\ref{fig2}
(lower subplot), resonators with a moderate $Q$ still can provide a
reasonable efficiency of generation in NDC-regime for SLs with the electron
densities $N=10^{15}-10^{16}$ cm$^{-3}$.
\par
We also found that with an increase of $\omega/\omega_d$ the value
of $\delta$ quickly decreases (see Inset in Fig.~\ref{fig3}).
Therefore, for higher frequencies or lower doping
($\omega_d\propto N$) providing $\omega/\omega_d>0.1$, the
influence of domains on the gain is practically negligible.
\par
We have made additionally the ensemble Monte Carlo simulations of space-charge dynamics \cite{ensemb_mc}.
These 3D simulations confirmed, in general, our main conclusions made in the framework of 1D DD-model.
Moreover, within the Monte Carlo approach we went beyond the LSA conditions considering
a short ($18$ periods) but heavy doped ($N=10^{17}$ cm$^{-3}$) SL.
In this case we found that $\delta$ is only $\approx 14$\% for $U_\omega=4U_c$.
\par
In practical terms, the SL oscillator with the parametric
cascading generation, i.e. the frequency conversion followed by
the parametric amplification, can operate in the whole microwave
and THz  frequency bands. In the present work we restrict our
treatment to the microwave and low-THz frequency range since there
exists plenty of serial radiation emitters which can provide a
necessary power to pump the SL oscillator. Probably the most
interesting and easily realizable case is to use a pump with
$\omega$ near 100 GHz in order to obtain a generation in the
important frequency band of hundreds of GHz \cite{pc}. In order to
suppress electrical domains and to use resonator with a reasonable
$Q$, the doping of SL should be $\simeq 10^{16}$ cm$^{-3}$. The
interaction of miniband electrons with the electric field for such
frequencies is still quasistatic.
\par
However, it is instructive to discuss briefly the effects arising
beyond the quasistatic approximation. Using the exact solution of
Boltzmann transport equation in the constant relaxation time
approximation \cite{chambers}, we have found the following two
effects. First, if $\omega$ is higher than some critical
$\omega_{cr}$, even in an ideal cavity a gain arises only if the
pump $E_{\omega}$ is larger than some threshold value $E_{th}$.
For the gain at third harmonic $\omega_{cr}\approx
0.34\tau^{-1}$($\approx 250$ GHz); importantly, we have
$E_{th}<E_c$ for the frequencies satisfying $\omega<0.38\tau^{-1}$($\approx
270$ GHz). Therefore, the generation of radiation at $3\omega$
without NDC and, as consequence without an influence of domains,
is possible for the pump field with frequencies $\omega<270$
GHz. The second effect, which can arise beyond the quasistatic
approximation, is the possibility to have a gain at the $5$th and the $9$th
harmonics. For example, a small gain at $5$th harmonic arises for
$\omega>0.14\tau^{-1}$($\approx 100$ GHz).
\par
Theoretical research devoted to a gain arising at frequency multiplication
in SLs has some history. To the best of our knowledge, Pavlovich
presented the first calculations showing a principle possibility
of a small-signal gain in the presence of a strong pump with
$\omega\tau\gtrsim 1$, if the frequency of the probe is some
half-integer of $\omega$ \cite{pavlovich77,remark_pavlovich}.
Romanov pointed out that such a gain should originate from a
multiphoton parametric amplification \cite{romanov80}. Possibility
to get a gain at harmonics of microwave pump also has been discussed
in \cite{renk_klapp,remark_renkklapp}. Because references to each
other are absent in the papers
\cite{pavlovich77,romanov80,renk_klapp}, we suppose that all these
works were done independently. This interesting activity did not
receive much attention so far. We should also underline that a
possible role of space-charge instabilities in SL has not been
analyzed in these previous contributions.
\par
Part of this work was done during stay of K.N.A in Loughborough
University; K.N.A. thanks Feo Kusmartsev for hospitality, fruitful
discussions and support. N.V.D. is grateful to the University of
Oulu for hospitality and Academy of Finland for support. We thank
Lauri Kurki for collaboration, Dmitry Pavelev and Dmitry Ryndyk
for the encouragement at earlier stages of the work,
Natalia Alexeeva, Timo Hyart, Aleksey Shorokhov, Feo Kusmartsev
and Erkki Thuneberg for critical reading of the manuscript.
K.-F. Renk shared our results during the whole 2004.
This research was supported by
Academy of Finland (grant 1206063) and  AQDJJ Programme of
European Science Foundation.

\appendix
\section{Appendix A. Weak-signal absorption in the quasistatic limit}
Here we derive the expression for absorption of the weak probe
field $E_{pr}=E_n\cos(n\omega t)$ in the presence of strong pump
field $E_{p}=E_{\omega}\cos\omega t$, ($E_n\ll E_0$).
\par
\begin{figure}[htbp!]
\includegraphics[width=1\linewidth]{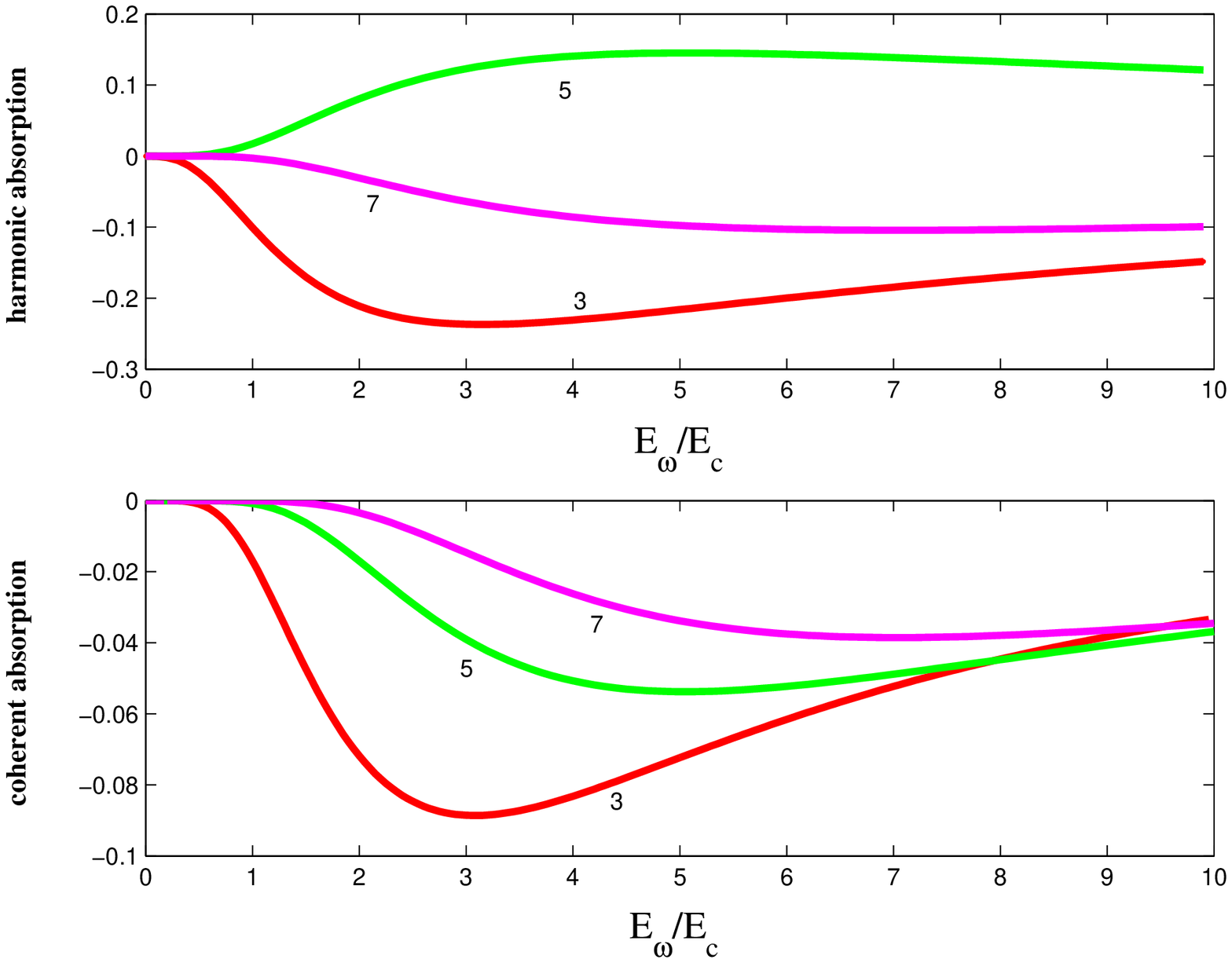}
\caption{ \label{fig_apend1} (Color) Small-signal harmonic $A_{h}$
[Eq. (\ref{a_harm_apend})] (upper subplot) and coherent
$A_{coh}/E_n$ [Eq. (\ref{a_coh_apend})] (lower subplot) components
of the absorption as a function of the pump field amplitude
$E_{\omega}/E_c$  and for different harmonics $n=3,5,7$.}
\end{figure}

Introduce the Fourier expansions for the electron velocity $V(E(t))$ and its derivative
$V^{\prime}(E(t))$ (prime means derivation with respect to $E$) as
\begin{equation}
\label{four_exp_def}
Y(E_{pump}(t))=\frac{1}{2}\tilde{Y}_0+\sum_{l=1}^{\infty}\tilde{Y}_l\cos(l\omega t),
\end{equation}
where $Y$ stands either for $V$ or for $V^{\prime}$.
Fourier transform $\tilde{Y}_l$ is defined as
\begin{equation}
\label{four_exp_coeff_def}
\tilde{Y}_l=\frac{2}{T}\int_0^T Y(E_{p}(t))\cos(l\omega t) dt
\end{equation}
Substituting the Fourier expansion (\ref{four_exp_def}) and the
Taylor expansion
$$
V(E)\approx V(E_{p})+V^{\prime}(E_{p})\times E_{pr}
$$
in the definition of absorption, we obtain
\begin{widetext}
$$
A=V_{p}^{-1}\langle V(E_{p})\cos(n\omega t)\rangle+
V_{p}^{-1}\langle V^{\prime}(E_{p})\cos^{2}(n\omega t)\rangle E_n
=
$$
$$
V_{p}^{-1} \left[
\sum_{l=1}^{\infty}\tilde{V}_l\langle\cos(l\omega t)\cos(n\omega
t)\rangle +E_n\left(
\sum_{l=1}^{\infty}\tilde{V}^{\prime}_l\langle\cos^{2}(n\omega
t)\cos(l\omega t)\rangle+
\frac{1}{2}\tilde{V}^{\prime}_0\langle\cos^{2}(n\omega t)\rangle
\right) \right]
$$
\begin{equation}
=\tilde{V}_n/(2V_{p})+\left[ \tilde{V}^{\prime}_0/V_{p}
+\tilde{V}^{\prime}_{2n}/V_{p} \right] E_n/4 \equiv
A_{harm}(E_{\omega})+A_{inc}(E_{\omega},E_n)+A_{coh}(E_{\omega},E_n).
\end{equation}
\end{widetext}
Using Esaki-Tsu characteristic, its derivative
\begin{equation}
V^{\prime}(E)=\frac{2 V_p}{E_c}\frac{1-{\bar E}^2}{\left(
1+{\bar E}^2\right)^2}, \quad {\bar E}=\frac{E}{E_c}
\end{equation}
and the definition of Fourier coefficients (\ref{four_exp_coeff_def}),
we calculate
\begin{equation}
\label{a_harm_apend} A_{h}\equiv
\frac{\tilde{V}_n}{2V_{p}}=(-1)^k \frac{2\left( b-1
\right)^{2k+1}}{b{\bar E}_{\omega}^{2k+1}},\quad b\equiv \left(
1+{\bar E}_{\omega}^2 \right)^{1/2}
\end{equation}
($2k+1=n$, $k=0,1,2,\ldots$, $n=3,5,7\ldots$; ${\bar E}_{\omega,n}\equiv E_{\omega,n}/E_c$)
for the harmonic component of total absorption
(see Fig. \ref{fig_apend1}, upper),
\begin{equation}
\label{a_inc_apend}
A_{inc}\equiv\frac{E_n}{4V_{p}}\tilde{V}^{\prime}_0=
\frac{{\bar E}_n}{\left( 1+{\bar E}_{\omega}^2\right)^{3/2}}
\end{equation}
for the incoherent component of total absorption, and
\begin{equation}
\label{a_coh_apend}
A_{coh}\equiv\frac{E_n}{4V_{p}}\tilde{V}^{\prime}_{2n}=
\frac{{\bar E}_n}{2\pi}\int_0^{2\pi} \frac{\left( 1-{\bar
E}_{\omega}^2\cos^2 x\right)\cos(2 n x)} {\left( 1+{\bar
E}_{\omega}^2\cos^2 x\right)^2}dx
\end{equation}
\begin{figure}[htbp!]
\includegraphics[width=0.7\linewidth]{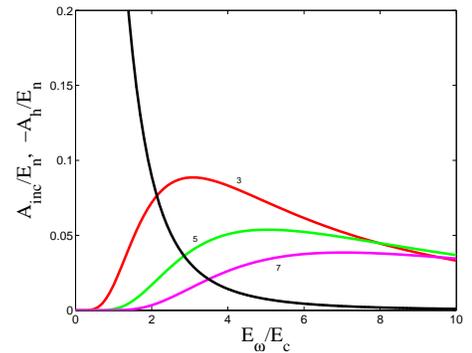}
\caption{ \label{fig_apend2} (Color) Comparison of the incoherent
and coherent absorption components. Dependencies of $A_{inc}/E_n$
(black curve) and $-A_{coh}/E_n$ (for different $n$) on  the pump
field amplitude $E_{\omega}/E_c$. }
\end{figure}
for the coherent component of total absorption.
\par
Note that $A_{inc}/E_n$, Eq.~(\ref{a_inc_apend}), is independent on $n$; it is always positive.
The integral (\ref{a_coh_apend}) can be calculated analytically in two limiting
cases ${\bar E}_{\omega}\ll 1$ and ${\bar E}_{\omega}\gg 1$. For
instance, for $n=3$ and ${\bar E}_{\omega}\ll 1$, we expand
$V^{\prime}(E)$ up to the term of ${\bar E}_{\omega}^7$ and get
$A_{coh}\approx -(7/64) {\bar E}_{\omega}^6{\bar E}_3$. In the
limit ${\bar E}_{\omega}\gg 1$ it easy to obtain $A_{coh}\approx
-(2n{\bar E}_n)/{\bar E}_{\omega}^2$ for any $n$. In general case
the integral (\ref{a_coh_apend}) can be easily calculated
numerically, see Fig.~\ref{fig_apend1}, lower. Note that $A_{coh}$ is always
negative.
\par
It is also instructive to compare $A_{inc}$ and $A_{coh}$ (see
Fig.~\ref{fig_apend2}).
For ${\bar E}_{\omega}\ll 1$ we have $|A_{coh}/A_{inc}|\simeq {\bar E}_{\omega}^{2n}
\ll 1$. In the opposite limit ${\bar E}_{\omega}\gg 1$, we have
$|A_{coh}/A_{inc}|\simeq {\bar E}_{\omega}\gg 1$. Thus, a contribution of the coherent
component to the total absorption is larger than a contribution of the incoherent
component for a large pump amplitude. Of course, a contribution of both
the coherent and the incoherent components is still less than a contribution of the
harmonic component.
\section{Appendix B. Dynamics of the field in the cavity}
Here we present the equations describing dynamics of electric
field in the cavity and derive formula for the SL oscillator
efficiency with an account of finite $Q$-factor of the cavity.
\par
For simplicity we consider a single-mode cavity with the
eigenfrequency $\omega_c$ tuned to the $n$th harmonic:
$\omega_c\approx\omega_n$. We search the field in the cavity
$E_{field}$ in the form
\begin{equation}
\label{field_form}
E_{field}=E_n(t)\cos[\omega_n t +\phi_n(t)],
\end{equation}
where the amplitude  $E_n(t)$ and the phase $\phi(t)$ are
slow-varying in comparison to the carrier frequency $\omega_n$,
i.e.
\begin{equation}
\label{SVEA_conditions} |\dot{E}_n|\ll\omega_n |E_n|, \quad
|\dot{\phi}_n|\ll\omega_n |\phi_n|.
\end{equation}
Substituting (\ref{field_form}) in the Maxwell equations and using
the so-called slow varying envelope approximation (SVEA), which is valid in
conditions (\ref{SVEA_conditions}), one can obtain \cite{lasers}
\begin{equation}
\label{cavity_1a}
\frac{\partial E_n}{\partial t}=-\frac{\gamma_c}{2}E_n
-\frac{2\pi}{\epsilon} J_c(k_n,t),
\end{equation}
\begin{equation}
\label{cavity_1b}
\frac{\partial\phi_n}{\partial t} E_n+(\omega_n-\omega_c) E_n=
-\frac{2\pi}{\epsilon} J_s(k_n,t),
\end{equation}
\begin{equation}
\label{cavity_1c}
J_{c,s}(k_n,t)=\frac{1}{L}\int_0^L dx R(x,k_n) j_{c,s},
\end{equation}
where $\gamma_c$ is the phenomenological loss constant, $R(x)$ is
the cavity mode with wavenumber $k_n$ ($R(x)\simeq\exp(i k_n x)$
and $\omega_c=c k_n$ for travelling wave mode; $R(x)\simeq\sin(
k_n x)$ for standing wave mode). In what follows we will neglect
spatial dependencies of $J_{c,s}$ because wavelength of microwave
or THz radiation is larger than nanostructure length, $k_n L\simeq
L/\lambda\ll 1$. Involved in Eq.~(\ref{cavity_1c}) cosine and sine
Fourier transformations of the current are defined as
$$
j_c=2 j_0 \langle v(t)\cos\omega_n t\rangle,\quad
j_s=2 j_0 \langle v(t)\sin\omega_n t\rangle,
$$
where $j_0=e V_p N$ and $v(t)\equiv V(t)/V_p$ is the scaled velocity of an
electron in SL arising
under the action of ac field $E(t)=E_p(t)+E_{field}(t)$.
\par
Next, we suppose that $\phi_n(0)=0$ and  the exact resonance,
$\omega_n=\omega_c$. Interaction of miniband electrons with the
field $E(t)$ is quasistatic and it is determined by the Esaki-Tsu
dependence $V(t)=V[E(t)]$. In these conditions we have
$j_s(t)=j_s(0)=0$, $\phi_n(t)=\phi_n(0)=0$ and $j_c=2 j_0 A$. This
approach means that for the quasistatic interactions of electrons
with fields we can neglect dispersion effects in the cavity. Now
we have single equation for the field in the cavity
\begin{equation}
\label{cavity_2}
\frac{\partial{\bar E}_n}{\partial
t}=-\frac{\gamma_c\tau}{2}{\bar E}_n -\frac{\omega^2_{pl}\tau}{2}
A(t)
\end{equation}
with $\omega^2_{pl}=4\pi e^2 N/(\epsilon m_0)$,
$m_0=(2\hbar^2)/(\Delta a^2)$ and ${\bar E}_n(t)=E_n(t)/E_c$.
Stationary solution of Eq.~(\ref{cavity_2}), ${\bar E}_n^{st}$,
has the form
\begin{equation}
\label{cavity_st}
\frac{{\bar E}_n^{st}}{Q}=-\frac{\omega^2_{pl}\tau}{\omega_n}
A\left( {\bar E}_{\omega},{\bar E}_n^{st}\right),
\end{equation}
where $Q=\omega_n/\gamma_c$. This formula has been used to plot
Fig.~\ref{fig3}.
\par
As follows from (\ref{cavity_2}), the stationary value of the field inside the
cavity is reached during the characteristic time
$\simeq\gamma_c$, i.e. $|\dot{E_n}/E_n|\simeq\gamma_c$. Therefore
the condition of applicability of SVEA (\ref{SVEA_conditions}) is
satisfied in a cavity with enough high $Q$:
$\gamma_c/\omega_n=1/Q\ll 1$.

\end{document}